\documentclass[ 
twocolumn,
prl, tightenlines,
nofootinbib,
amsfonts,amsmath,amssymb]{revtex4}
\usepackage{bm} \usepackage{graphics,graphicx,epsfig,color}
\usepackage{subfigure} \usepackage{hyperref}

\allowdisplaybreaks
\begin{document}
 
\title{Gravit\'e modifi\'ee ou mati\`ere modifi\'ee~?\,\footnote{A para\^itre
    dans la revue \textit{l'Astronomie} (2009).}}

\author{Luc \textsc{Blanchet}} 
%\email{blanchet@iap.fr} 
\affiliation{Institut d'Astrophysique de Paris --- CNRS, Universit\'e Pierre
  \& Marie Curie}

%\date{\today}

\begin{abstract}
Pour l'astrophysicien qui aborde le puzzle de la mati\`ere noire, celle-ci
appara\^it sous deux aspects diff\'erents: d'une part en cosmologie,
c'est-\`a-dire \`a tr\`es grandes \'echelles, o\`u elle semble \^etre form\'ee
d'un bain de particules, et d'autre part \`a l'\'echelle des galaxies, o\`u
elle est d\'ecrite par un ensemble de ph\'enom\`enes tr\`es particuliers, qui
paraissent incompatibles avec sa description en termes de particules, et qui
font dire \`a certains que l'on est en pr\'esence d'une modification de la loi
de la gravitation. R\'econcilier ces deux aspects distincts de la mati\`ere
noire dans un m\^eme formalisme th\'eorique repr\'esente un d\'efi important
qui pourrait peut-\^etre conduire \`a une physique nouvelle en action aux
\'echelles astronomiques.
\end{abstract}

%\pacs{95.35.+d,95.36.+x,04.50.Kd}

\maketitle

\section{Mati\`ere noire en cosmologie}

Dans le mod\`ele cosmologique, dit de concordance car il est en conformit\'e
avec tout un ensemble de donn\'ees observationnelles, la mati\`ere ordinaire
dont sont constitu\'es les \'etoiles, le gaz, les galaxies,
etc. (essentiellement sous forme baryonique) ne forme que 4\% de la
masse-\'energie totale, ce qui est d\'eduit de la nucl\'eosynth\`ese
primordiale des \'el\'ements l\'egers, ainsi que des mesures de fluctuations
du rayonnement du fond diffus cosmologique (CMB) --- le rayonnement fossile
qui date de la formation des premiers atomes neutres dans l'Univers. Nous
savons aussi qu'il y a 23\% de mati\`ere noire sous forme \textit{non
  baryonique} et dont nous ne connaissons pas la nature. Et les 73\% qui
restent~?  Et bien, ils sont sous la forme d'une myst\'erieuse \'energie
noire, mise en \'evidence par le diagramme de Hubble des supernovae de type
Ia, et dont on ignore l'origine \`a part qu'elle pourrait \^etre sous la forme
d'une constante cosmologique. Le contenu de l'univers \`a grandes \'echelles
est donc donn\'e par le ``camembert'' de la figure \ref{fig1} dont 96\% nous
est inconnu~!
\begin{figure}[h]
\centering{\includegraphics[width=3in]{fig1.jpg}
\caption{Le contenu en masse-\'energie de l'Univers.}
\label{fig1}}
\end{figure}

\subsection{Un mod\`ele \`a succ\`es}

La mati\`ere noire permet d'expliquer la diff\'erence entre la masse dynamique
des amas de galaxies (c'est la masse d\'eduite du mouvement des galaxies) et
la masse de la mati\`ere lumineuse qui comprend les galaxies et le gaz chaud
intergalactique. Mais cette mati\`ere noire ne fait pas que cela~! Nous pensons
qu'elle joue un r\^ole crucial dans la formation des grandes structures, en
entra\^inant la mati\`ere ordinaire dans un effondrement gravitationnel, ce
qui permet d'expliquer la distribution de mati\`ere visible depuis l'\'echelle
des amas de galaxies jusqu'\`a l'\'echelle cosmologique. Des simulations
num\'eriques tr\`es pr\'ecises permettent de confirmer cette hypoth\`ese. Pour
que cela soit possible il faut que la mati\`ere noire soit non relativiste au
moment de la formation des galaxies. On l'appelera mati\`ere noire
\textit{froide} ou CDM selon l'acronyme anglais, et il y a aussi un nom pour
la particule associ\'ee: un WIMP pour ``weakly interacting massive particle''.

Il n'y a pas d'explication pour la mati\`ere noire (ni pour l'\'energie noire)
dans le cadre du mod\`ele standard de la physique des particules. Mais des
extensions au-del\`a du mod\`ele standard permettent de trouver des bons
candidats pour la particule \'eventuelle de mati\`ere noire. Par exemple dans
un mod\`ele de super-sym\'etrie (qui associe \`a tout fermion un partenaire
super-sym\'etrique qui est un boson et r\'eciproquement) l'un des meilleurs
candidats est le \textit{neutralino}, qui est un partenaire fermionique
super-sym\'etrique d'une certaine combinaison de bosons du mod\`ele
standard. L'\textit{axion}, qui fut introduit dans une tentative pour
r\'esoudre le probl\`eme de la violation CP en physique des particules, est
une autre possibilit\'e. Il y a aussi les \'etats de Kaluza-Klein pr\'edits
dans certains mod\`eles avec dimensions supl\'ementaires.

Quant \`a l'\'energie noire, elle appara\^it comme un milieu de densit\'e
d'\'energie \textit{constante} au cours de l'expansion, ce qui implique une
violation des ``conditions d'\'energie'' habituelles avec une pression
n\'egative. L'\'energie noire pourrait \^etre la fameuse constante
cosmologique $\Lambda$ qu'Einstein avait introduite dans les \'equations de la
relativit\'e g\'en\'erale afin d'obtenir un mod\`ele d'univers statique, puis
qu'il avait abandonn\'ee lorsque l'expansion fut d\'ecouverte. Depuis
Zel'dovich on interpr\`ete $\Lambda$ comme l'\'energie du vide associ\'ee \`a
l'espace-temps lui-m\^eme. Le probl\`eme est que l'estimation de cette
\'energie en th\'eorie des champs donne une valeur $10^{123}$ fois plus grande
que la valeur observ\'ee~! On ne comprend donc pas pourquoi la constante
cosmologique est si petite.

Malgr\'e l'\'enigme de l'origine de ses constituents, le mod\`ele
$\Lambda$-CDM est plein de succ\`es, tant dans l'ajustement pr\'ecis des
fluctuations du CMB que dans la reproduction fid\`ele des grandes structures
observ\'ees. Une le\c{c}on est que la mati\`ere noire appara\^it form\'ee de
particules (les WIMPs) \`a grande \'echelle.

\section{Mati\`ere noire dans les galaxies}

\begin{figure}[h]
\centering{\includegraphics[width=3.2in]{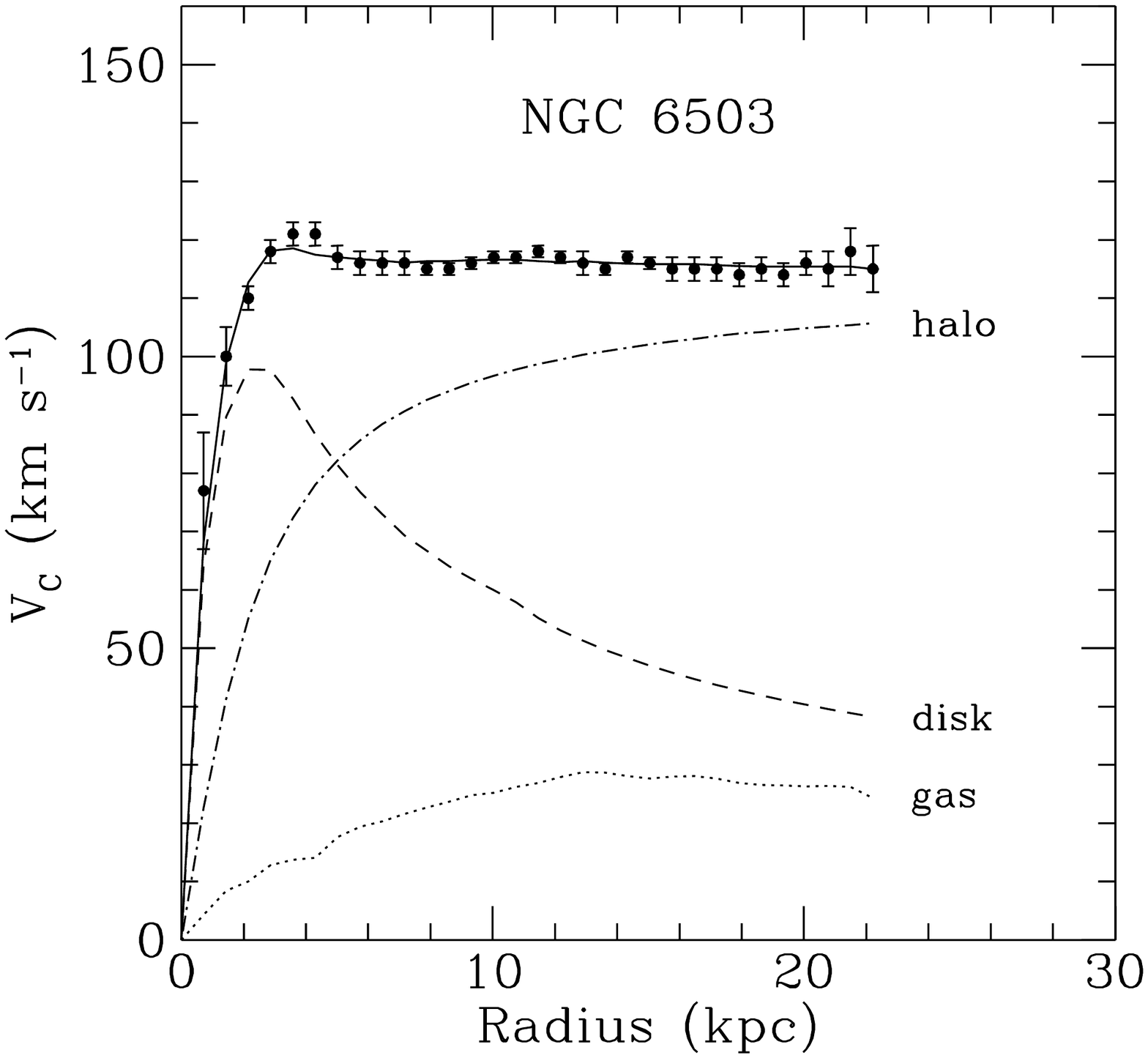}
\vspace{-2.3cm}
\caption{Courbe de rotation de galaxie.}
\label{fig2}}
\end{figure}

La mati\`ere noire se manifeste de mani\`ere \'eclatante dans les galaxies,
par l'exc\`es de vitesse de rotation des \'etoiles autour de ces galaxies en
fonction de la distance au centre --- c'est la c\'el\`ebre courbe de rotation
(voir la figure \ref{fig2}). Les mesures montrent qu'\`a partir d'une certaine
distance au centre la courbe de rotation devient pratiquement plate,
c'est-\`a-dire que la vitesse devient constante.
\vspace{-1.9cm}

D'apr\`es la loi de Newton la vitesse d'une \'etoile sur une orbite circulaire
(keplerienne) de rayon $r$ est donn\'ee par $v(r)=\sqrt{G M(r)/r}$ o\`u $M(r)$
est la masse contenue dans la sph\`ere de rayon $r$. Pour obtenir une courbe
de rotation plate il faut donc supposer que la masse cro\^it
proportionnellement \`a $r$ (et donc que la densit\'e d\'ecro\^it comme
$1/r^2$), ce qui n'est certainement pas le cas de la mati\`ere visible. On est
oblig\'e d'invoquer l'existence d'un gigantesque halo de mati\`ere noire
invisible (qui ne rayonne pas) autour de la galaxie et dont la masse
dominerait celle des \'etoiles et du gaz.

\subsection{Probl\`emes dans les halos}

Cette mati\`ere noire peut-elle \^etre faite de la m\^eme particule que celle
sugg\'er\'ee par la cosmologie (un WIMP)~? Des \'el\'ements de r\'eponse sont
fournis par les simulations num\'eriques de CDM en cosmologie qui sont aussi
valables \`a l'\'echelle des galaxies, et qui donnent un profil de densit\'e
universel pour le halo de mati\`ere noire. A grande distance ce profil
d\'ecroit en $1/r^3$ soit plus rapidement que ce qu'il faudrait pour avoir une
courbe plate, mais ce n'est pas tr\`es grave car on peut supposer que la
courbe de rotation est observ\'ee dans un r\'egime interm\'ediaire avant de
d\'ecro\^itre. Plus grave est la pr\'ediction d'un pic central de densit\'e au
centre des galaxies, o\`u les particules de mati\`ere noire tendent \`a
s'agglom\'erer \`a cause de la gravitation, avec une loi en $1/r$ pour $r$
petit. Or les courbes de rotation favorisent plut\^ot un profil de densit\'e
sans divergence, avec un coeur de densit\'e constante.

\begin{figure}[h]
\centering{\includegraphics[width=3.5in]{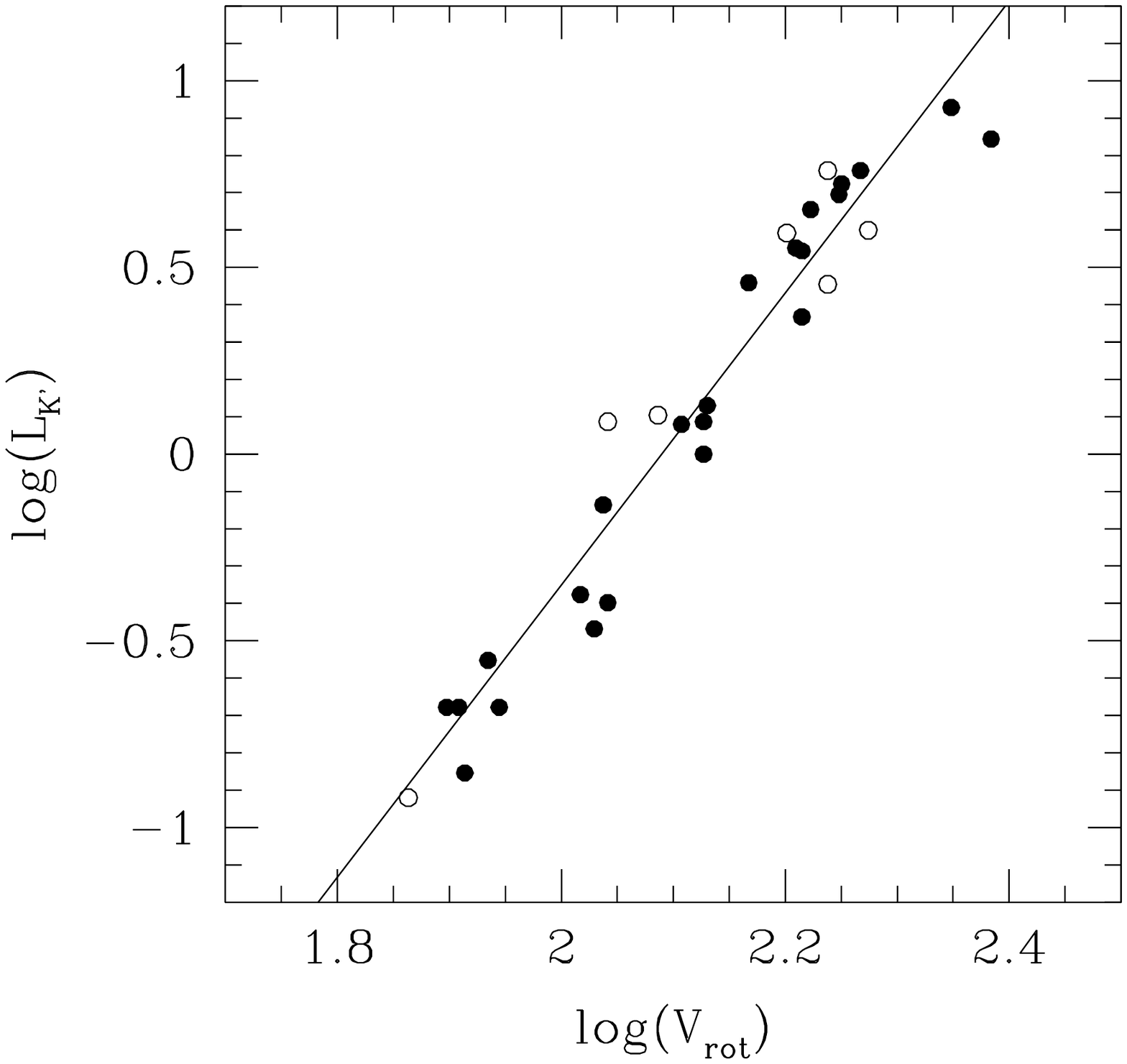}
\vspace{-3.3cm}
\caption{Loi de Tully-Fisher.}
\label{fig3}}
\end{figure}

D'autres probl\`emes rencontr\'es par les halos simul\'es de CDM sont la
formation d'une multitude de satellites autour des grosses galaxies, et la loi
empirique de Tully et Fisher qui n'est pas expliqu\'ee de fa\c{c}on
naturelle. Cette loi montr\'ee dans la figure \ref{fig3} relie la luminosit\'e
des galaxies \`a leur vitesse asymptotique de rotation (qui est la valeur du
plateau dans la figure \ref{fig2}) par $v\propto L^{1/4}$. Noter que cette loi
ne fait pas r\'ef\'erence \`a la mati\`ere noire~! La vitesse et la
luminosit\'e sont bien s\^ur celles de la mati\`ere ordinaire, et la mati\`ere
noire semble faire ce que lui dicte la mati\`ere visible.
\vspace{-1.7cm}

Mais le d\'efi le plus important de CDM est de pouvoir rendre compte d'une
observation \'etonnante appel\'ee \textit{loi de Milgrom} \cite{Mil}, selon
laquelle la mati\`ere noire intervient uniquement dans les r\'egions o\`u le
champ de gravitation (ou, ce qui revient au m\^eme, le champ
d'acc\'el\'eration) est plus \textit{faible} qu'une certaine acc\'el\'eration
critique mesur\'ee \`a la valeur ``universelle'' $a_0\simeq 1,2\times
10^{-10}\,\mathrm{m}/\mathrm{s}^2$. Tout se passe comme si dans le r\'egime
des champs faibles $g\ll a_0$, la mati\`ere ordinaire \'etait acc\'el\'er\'ee
non par le champ newtonien $g_\text{N}$ mais par un champ $g$ donn\'e
simplement par $g=(a_0 g_\text{N})^{1/2}$. La loi du mouvement sur une orbite
circulaire donne alors une vitesse \textit{constante} et \'egale \`a $v=(G M
a_0)^{1/4}$. Ce r\'esultat nous r\'eserve un bonus important: puisque le
rapport masse-sur-luminosit\'e $M/L$ est approximativement le m\^eme d'une
galaxie \`a l'autre, la vitesse de rotation doit varier comme la puissance
$1/4$ de la luminosit\'e $L$, en accord avec la loi de Tully-Fisher~!

\subsection{Une formule puissante}

Pour avoir une r\`egle qui nous permette d'ajuster les courbes de rotation des
galaxies il nous faut aussi prendre en compte le r\'egime de champ fort dans
lequel on doit retrouver la loi newtonienne. On introduit une fonction
d'interpolation $\mu$ d\'ependant du rapport $g/a_0$ et qui se ram\`ene \`a
$\mu\simeq g/a_0$ lorsque $g\ll a_0$, et qui tend vers 1 quand $g\gg
a_0$. Notre r\`egle sera donc
\begin{equation}\label{mond0}
	\mu\left(g/a_0\right) \bm{g} = \bm{g}_\text{N} \,.
\end{equation}
Ici $g$ d\'esigne la norme du champ de gravitation $\bm{g}$ ressenti par les
particules d'\'epreuves. Une formule encore plus op\'erationnelle est obtenue
en prenant la divergence des deux membres de \eqref{mond0} ce qui m\`ene \`a
l'\'equation de Poisson modifi\'ee\footnote{L'\'equation de Poisson usuelle
  s'\'ecrit: $\Delta U =-4 \pi \, G\,\rho_\text{b}$, o\`u $\Delta$ est le
  laplacien et $U$ le potentiel newtonien local. L'op\'erateur $\bm{\nabla}$
  appliqu\'e \`a une fonction scalaire est le gradient, appliqu\'e \`a un
  vecteur c'est la divergence: $\Delta U=\bm{\nabla}\cdot\bm{\nabla}U$. Par
  convention, on note les vecteurs en caract\`eres gras.}:
\begin{equation}\label{mond}
	\bm{\nabla} \! \cdot \Bigl[ \mu\left(g/a_0\right) \bm{g} \Bigr] = -4
        \pi \, G\,\rho_\text{b} \,,
\end{equation}
dont la source est la densit\'e de mati\`ere baryonique $\rho_\text{b}$ (le
champ gravitationnel est irrotationnel: $\bm{g}=\bm{\nabla}U$). On appellera
l'\'equation \eqref{mond} la formule MOND pour ``modified Newtonian
dynamics''.

Le succ\`es de cette formule (on devrait plus exactement dire cette
\textit{recette}) dans l'obtention des courbes de rotation de nombreuses
galaxies est impressionnant; voir la courbe en trait plein dans la figure
\ref{fig2}. C'est en fait un ajustement \`a un param\`etre libre, le rapport
$M/L$ de la galaxie qui est donc \textit{mesur\'e} par notre recette. On
trouve que non seulement la valeur de $M/L$ est de l'ordre de 1-5 comme il se
doit, mais qu'elle est remarquablement en accord avec la couleur observ\'ee de
la galaxie.

Beaucoup consid\`erent la formule MOND comme ``exotique'' et repr\'esentant un
aspect mineur du probl\`eme de la mati\`ere noire. On entend m\^eme parfois
dire que ce n'est pas de la physique. Bien s\^ur ce n'est pas de la physique
\textit{fondamentale} --- cette formule ne peut pas \^etre consid\'er\'ee
comme une th\'eorie fondamentale, mais elle constitue de l'excellente
physique~! Elle capture de fa\c{c}on simple et puissante tout un ensemble de
faits observationnels. Au physicien th\'eoricien d'expliquer pourquoi.

La valeur num\'erique de $a_0$ se trouve \^etre tr\`es proche de la constante
cosmologique: $a_0\sim c^2\sqrt{\Lambda}$. Cette co\"incidence cosmique
pourrait nous fournir un indice~! Elle a aliment\'e de nombreuses
sp\'eculations sur une possible influence de la cosmologie dans la dynamique
locale des galaxies.

Face \`a la ``d\'eraisonnable efficacit\'e'' de la formule MOND, trois
solutions sont possibles.
\begin{enumerate}
\item La formule pourrait s'expliquer dans le cadre CDM. Mais pour r\'esoudre
  les probl\`emes de CDM il faut invoquer des m\'ecanismes astrophysiques
  compliqu\'es et effectuer un ajustement fin des donn\'ees galaxie par
  galaxie.
\item On est en pr\'esence d'une modification de la loi de la gravitation dans
  un r\'egime de champ faible $g\ll a_0$. C'est l'approche traditionnelle de
  MOND et de ses extensions relativistes.
\item La gravitation n'est pas modifi\'ee mais la mati\`ere noire poss\`ede
  des caract\'eristiques particuli\`eres la rendant apte \`a expliquer la
  ph\'enom\'enologie de MOND. C'est une approche nouvelle qui se pr\^ete aussi
  tr\`es bien \`a la cosmologie.
\end{enumerate}
La plupart des astrophysiciens des particules et des cosmologues des grandes
structures sont partisans de la premi\`ere solution. Malheureusement aucun
m\'ecanisme convainquant n'a \'et\'e trouv\'e pour incorporer de fa\c{c}on
naturelle la constante d'acc\'el\'eration $a_0$ dans les halos de CDM. Dans la
suite nous consid\'ererons que la solution 1. est d'ores et d\'ej\`a exclue par
les observations.

Les approches 2. de gravitation modifi\'ee et 3. que l'on peut qualifier de
\textit{mati\`ere noire modifi\'ee} croient toutes deux dans la pertinence de
MOND, mais comme on va le voir sont en fait tr\`es diff\'erentes. Notez que
dans ces deux approches il faudra expliquer pourquoi la mati\`ere noire semble
\^etre constitu\'ee de WIMPs \`a l'\'echelle cosmologique.

\section{Th\'eories de la gravitation modifi\'ee}

Cette route, tr\`es d\'evelopp\'ee dans la litt\'erature, consiste \`a
supposer qu'il n'y a pas de mati\`ere noire, et que
\eqref{mond0}--\eqref{mond} refl\`ete une violation fondamentale de la loi de
la gravitation. C'est la proposition initiale de Milgrom \cite{Mil} --- un
changement radical de paradigme par rapport \`a l'approche CDM. Pour esp\'erer
d\'efinir une th\'eorie il nous faut partir d'un lagrangien\footnote{Le
  lagrangien est une fonction des variables dynamiques d'un syst\`eme
  (positions et vitesses, en m\'ecanique classique) qui permet de d\'ecrire de
  mani\`ere concise les \'equations du mouvement du syst\`eme. Ces derni\`eres
  s'obtiennent par application du principe de moindre action (ou principe
  d'action extr\'emale).}. Or il est facile de voir que \eqref{mond} d\'ecoule
d'un lagrangien, celui-ci ayant la particularit\'e de comporter un terme
cin\'etique non standard pour le potentiel gravitationnel, du type
$f[(\bm{\nabla}U)^2]$ au lieu du terme habituel $(\bm{\nabla}U)^2$, o\`u $f$
est une certaine fonction que l'on relie \`a la fonction $\mu$. Ce lagrangien
a servi de point de d\'epart pour la construction des th\'eories de la
gravitation modifi\'ee.

On veut modifier la relativit\'e g\'en\'erale de fa\c{c}on \`a retrouver MOND
dans la limite non-relativiste, c'est-\`a-dire quand la vitesse des corps est
tr\`es faible par rapport \`a la vitesse de la lumi\`ere $c$. En relativit\'e
g\'en\'erale la gravitation est d\'ecrite par un champ tensoriel \`a deux
indices appel\'e la m\'etrique de l'espace-temps $g_{\alpha\beta}$. Cette
th\'eorie est extr\^emement bien v\'erifi\'ee dans le Syst\`eme Solaire et
dans les pulsars binaires, mais peu test\'ee dans le r\'egime de champs
faibles qui nous int\'eresse (en fait la relativit\'e g\'en\'erale est le
royaume des champs gravitationnels forts).

\subsection{Une premi\`ere tentative}

La premi\`ere id\'ee qui vient \`a l'esprit est de promouvoir le potentiel
newtonien $U$ en un champ scalaire $\phi$ (sans indices) et donc de
consid\'erer une th\'eorie \textit{tenseur-scalaire} dans laquelle la
gravitation est d\'ecrite par le couple de champs $(g_{\alpha\beta},\phi)$. On
postule, de mani\`ere \textit{ad-hoc}, un terme cin\'etique non standard pour
le champ scalaire: $F(\partial_\alpha\phi\partial^\alpha\phi)$ o\`u $F$ est
reli\'e \`a $\mu$, et on choisit le lagrangien d'Einstein-Hilbert de la
relativit\'e g\'en\'erale pour la partie concernant la m\'etrique
$g_{\alpha\beta}$.

Tout va bien pour ce qui concerne le mouvement des \'etoiles dans une galaxie,
qui reproduit MOND. Mais notre th\'eorie tenseur-scalaire est une catastrophe
pour le mouvement des photons~! En effet ceux-ci ne ressentent pas la
pr\'esence du champ scalaire $\phi$ cens\'e repr\'esenter la mati\`ere
noire. Dans une th\'eorie tenseur-scalaire toutes les formes de mati\`ere se
propagent dans un espace-temps de m\'etrique \textit{physique}
$\tilde{g}_{\alpha\beta}$ qui diff\`ere de la m\'etrique d'Einstein
$g_{\alpha\beta}$ par un facteur de proportionalit\'e d\'ependant du champ
scalaire, soit $\tilde{g}_{\alpha\beta}=A(\phi)g_{\alpha\beta}$. Une telle
relation entre les m\'etriques est dite conforme et laisse invariants les
c\^ones de lumi\`ere de l'espace-temps. Les trajectoires de photons seront
donc les m\^emes dans l'espace-temps physique que dans l'espace-temps
d'Einstein (cela se d\'eduit aussi de l'invariance conforme des \'equations de
Maxwell). Comme on observe d'\'enormes quantit\'es de mati\`ere noire gr\^ace
au mouvement des photons, par effet de lentille gravitationnelle, la th\'eorie
tenseur-scalaire est \'elimin\'ee.

\subsection{Th\'eorie tenseur-vecteur-scalaire}

Pour corriger cet effet d\'esastreux du mouvement de la lumi\`ere on rajoute
un nouvel \'el\'ement \`a notre th\'eorie. Puisque c'est cela qui cause
probl\`eme on va transformer la relation entre les m\'etriques
$\tilde{g}_{\alpha\beta}$ et $g_{\alpha\beta}$. Une fa\c{c}on de le faire est
d'y ins\'erer (encore de fa\c{c}on \textit{ad-hoc}) un nouveau champ qui sera
cette fois un vecteur $V^\alpha$ avec un indice. On aboutit donc \`a une
th\'eorie dans laquelle la gravitation est d\'ecrite par le triplet de champs
$(g_{\alpha\beta},V^\alpha,\phi)$. C'est ce qu'on appelle une th\'eorie
\textit{tenseur-vecteur-scalaire} (TeVeS).

La th\'eorie TeVeS a \'et\'e mise au point par Bekenstein et Sanders
\cite{Be,Sa}. Comme dans la th\'eorie tenseur-scalaire on aura la partie
d'Einstein-Hilbert pour la m\'etrique, plus un terme cin\'etique non standard
$F(\partial_\alpha\phi\partial^\alpha\phi)$ pour le champ scalaire. Quant au
champ vectoriel on le munit d'un terme cin\'etique analogue \`a celui de
l'\'electromagn\'etisme, mais dans lequel le r\^ole du potentiel
\'electromagn\'etique $A^\alpha$ est tenu par notre champ $V^\alpha$. La
th\'eorie TeVeS r\'esultante est tr\`es compliqu\'ee et pour l'instant non
reli\'ee \`a de la physique microscopique. Il a \'et\'e montr\'e que c'est un
cas particulier d'une classe de th\'eories appel\'ees th\'eories
Einstein-\textit{\'ether} dans lesquelles le vecteur $V^\alpha$ joue le r\^ole
principal, en d\'efinissant un r\'ef\'erentiel priviligi\'e un peu analogue
\`a l'\'ether postul\'e au XIX$^\text{\`eme}$ si\`ecle pour interpr\'eter la
non-invariance des \'equations de Maxwell par transformation de Galil\'ee.

\subsection{Les neutrinos \`a la rescousse}

Si elle est capable de retrouver MOND dans les galaxies, la th\'eorie TeVeS a
malheureusement un probl\`eme dans les amas de galaxies car elle ne rend pas
compte de toute la mati\`ere noire observ\'ee. C'est en fait un probl\`eme
g\'en\'erique de toute extension relativiste de MOND. Cependant ce probl\`eme
peut \^etre r\'esolu en supposant l'existence d'une composante de mati\`ere
noire \textit{chaude} sous la forme de neutrinos massifs, ayant la masse
maximale permise par les exp\'eriences actuelles soit environ
$2\,\text{eV}$. Rappelons que toute la mati\`ere noire ne peut pas \^etre sous
forme de neutrinos: d'une part il n'y aurait pas assez de masse, et d'autre
part les neutrinos \'etant relativistes auraient tendance \`a lisser
l'apparence des grandes structures, ce qui n'est pas observ\'e. N\'eanmoins
une pinc\'ee de neutrinos massifs pourrait permettre de rendre viables les
th\'eories de gravitation modifi\'ee. De ce point de vue les exp\'eriences
pr\'evues qui vont d\'eterminer tr\`es pr\'ecis\'ement la masse du neutrino
(en v\'erifiant la conservation de l'\'energie au cours de la
d\'esint\'egration d'une particule produisant un neutrino dans l'\'etat final)
vont jouer un r\^ole important en cosmologie. TeVeS a aussi des difficult\'es
\`a l'\'echelle cosmologique pour reproduire les fluctuations observ\'ees du
CMB. L\`a aussi une composante de neutrinos massifs peut aider, mais la
hauteur du troisi\`eme pic de fluctuation, qui est caract\'eristique de la
pr\'esence de mati\`ere noire sans pression, reste difficile \`a ajuster.

\section{Th\'eorie de la mati\`ere modifi\'ee}

Une alternative logique \`a la gravit\'e modifi\'ee est de supposer qu'on est
en pr\'esence d'une forme particuli\`ere de mati\`ere noire ayant des
caract\'eristiques diff\'erentes de CDM. Dans cette approche on a l'ambition
d'expliquer la ph\'enom\'enologie de MOND, mais avec une philosophie nouvelle
puisqu'on ne modifie pas la loi de la gravitation: on garde la relativit\'e
g\'en\'erale classique, avec sa limite newtonienne habituelle. Cette
possibilit\'e \'emerge gr\^ace \`a l'analogue gravitationnel du m\'ecanisme
physique de polarisation par un champ ext\'erieur et qu'on va appeler
``polarisation gravitationnelle'' \cite{B}.

\subsection{Une interpr\'etation de MOND}

La motivation physique est une analogie frappante (et peut-\^etre tr\`es
profonde) entre MOND, sous la forme de l'\'equation de Poisson modifi\'ee
\eqref{mond}, et la physique des milieux di\'electriques en
\'electrostatique. En effet nous apprenons dans nos cours de physique
\'el\'ementaire que l'\'equation de Gauss pour le champ \'electrique (c'est
l'une des \'equations fondamentales de Maxwell), est modifi\'ee en pr\'esence
d'un milieu di\'electrique par la contribution de la polarisation \'electrique
(voir l'Appendice). 

De m\^eme, MOND peut-\^etre vu comme la modification de l'\'equation de
Poisson par un milieu ``digravitationnel''. Explicitons cette analogie. On
introduit l'analogue gravitationnel de la susceptibilit\'e, soit $\chi$ qui
est reli\'e \`a la fonction MOND par $\mu=1+\chi$. La ``polarisation
gravitationnelle'' est d\'efinie par
\begin{equation}\label{Pig}
	\bm{\Pi} = -\frac{\chi}{4 \pi \, G}\,\bm{g}\, .
\end{equation}
La densit\'e des ``masses de polarisation'' est donn\'ee par la divergence de
la polarisation soit $\rho_\text{pol}=-\bm{\nabla} \!  \cdot \!
\bm{\Pi}$. Avec ces notations l'\'equation \eqref{mond} devient
\begin{equation}\label{mondpol}
	\bm{\nabla} \! \cdot \! \bm{g} = -4 \pi \, G\,\bigl(\rho_\text{b} +
        \rho_\text{pol}\bigr) \, ,
\end{equation}
qui appara\^it maintenant comme une \'equation de Poisson ordinaire, mais dont
la source est constitu\'ee non seulement par la densit\'e de mati\`ere
baryonique, mais aussi par la contribution des masses de polarisation
$\rho_\text{pol}$. Il est clair que cette \'ecriture de MOND sugg\`ere que
l'on est en pr\'esence non pas d'une modification de la loi gravitationnelle,
mais d'une forme nouvelle de mati\`ere noire de densit\'e $\rho_\text{pol}$,
c'est-\`a-dire faite de moments dipolaires align\'es dans le champ de
gravitation.

\subsection{Des masses n\'egatives~?}

L'\'etape suivante serait de construire un mod\`ele microscopique pour des
dip\^oles gravitationnels $\bm{\pi}$ (tels que
$\bm{\Pi}=n\bm{\pi}$). L'analogue gravitationnel du dip\^ole \'electrique
serait un vecteur $\bm{\pi}=m\bm{\xi}$ s\'eparant deux masses $\pm m$. On se
heurte donc \`a un probl\`eme s\'ev\`ere: le milieu dipolaire gravitationnel
devrait contenir des masses n\'egatives~! Ici on entend par masse l'analogue
gravitationnel de la charge, qui est ce qu'on appelle parfois la masse
grave. Ce probl\`eme des masses n\'egatives rend \textit{a priori} le mod\`ele
hautement non viable. N\'eanmoins, ce mod\`ele est int\'eressant car il est
facile de montrer que le coefficient de susceptibilit\'e gravitationnelle doit
\^etre n\'egatif, $\chi<0$, soit l'oppos\'e du cas \'electrostatique. Or c'est
pr\'ecis\'ement ce que nous dit MOND: comme la fonction $\mu$ interpole entre
le r\'egime MOND o\`u $\mu\ll 1$ et le r\'egime newtonien o\`u $\mu\rightarrow
1$, on a $\mu<1$ et donc bien $\chi<0$. Il est donc tentant d'interpr\'eter le
champ gravitationnel plus intense dans MOND que chez Newton par la pr\'esence
de ``masses de polarisation'' qui \textit{anti-\'ecrantent} le champ des
masses gravitationnel ordinaires, et ainsi augmentent l'intensit\'e effective
du champ gravitationnel~!

\subsection{Une cinqui\`eme force}

Dans le cadre de ce mod\`ele on peut aussi se convaincre qu'un milieu form\'e
de dip\^oles gravitationnels est intrins\`equement instable, car les
constituants microscopiques du dip\^ole devraient se repousser
gravitationnellement. Il faut donc introduire une force interne d'origine
\textit{non-gravitationnelle}, qui va supplanter la force gravitationnelle
pour lier les constituants dipolaires entre eux. On pourrait qualifier cette
nouvelle interaction de ``cinqui\`eme force''. Pour retrouver MOND, on trouve
de fa\c{c}on satisfaisante que ladite force doit d\'ependre du champ de
polarisation, et avoir en premi\`ere approximation la forme d'un oscillateur
harmonique. Par l'effet de cette force, \`a l'\'equilibre, le milieu dipolaire
ressemble \`a une sorte d'``\'ether statique'', un peu \`a l'image du
di\'electrique dont les sites atomiques sont fixes.

\subsection{Un mod\`ele relativiste}

Les arguments pr\'ec\'edents nous laissent penser que MOND a quelque chose \`a
voir avec un effet de polarisation gravitationnelle. Mais il nous faut
maintenant construire un mod\`ele coh\'erent, reproduisant l'essentiel de cette
physique, et \textit{sans} masses graves n\'egatives, donc respectant le
principe d'\'equivalence. Il faut aussi bien s\^ur que le mod\`ele soit
\textit{relativiste} (en relativit\'e g\'en\'erale) pour pouvoir r\'epondre
\`a des questions concernant la cosmologie ou le mouvement de photons.

On va d\'ecrire le milieu comme un fluide relativiste de quadri-courant
$J^\alpha=\rho u^\alpha$ (o\`u $\rho$ est la densit\'e de masse), et muni d'un
quadri-vecteur $\xi^\alpha$ jouant le r\^ole du moment dipolaire. Le vecteur
de polarisation est alors $\Pi^\alpha=\rho\xi^\alpha$. On d\'efinit un
principe d'action pour cette mati\`ere dipolaire, que l'on rajoute \`a
l'action d'Einstein-Hilbert, et \`a la somme des actions de tous les champs de
mati\`ere habituels (baryons, photons, etc). On inclue dans l'action une
fonction potentielle d\'ependant de la polarisation et cens\'ee d\'ecrire une
force interne au milieu dipolaire. Par variation de l'action on obtient
l'\'equation du mouvement du fluide dipolaire, ainsi que l'\'equation
d'\'evolution de son moment dipolaire. On trouve que le mouvement du fluide
est affect\'e par la force interne, et diff\`ere du mouvement g\'eod\'esique
d'un fluide ordinaire.

\subsection{Une bonne physique}

Ce mod\`ele (propos\'e dans \cite{BL}) reproduit bien la ph\'enom\'enologie de
MOND au niveau des galaxies. Il a \'et\'e construit pour~! Mais il a \'et\'e aussi
d\'emontr\'e qu'il donne satisfaction en cosmologie o\`u l'on consid\`ere une
perturbation d'un univers homog\`ene et isotrope. En effet cette mati\`ere
noire dipolaire se conduit comme un fluide parfait sans pression au premier
ordre de perturbation cosmologique et est donc indistinguable du mod\`ele
CDM. En particulier le mod\`ele est en accord avec les fluctuations du fond
diffus cosmologique (CMB). En ce sens il permet de r\'econcilier l'aspect
particulaire de la mati\`ere noire telle qu'elle est d\'etect\'ee en
cosmologie avec son aspect ``modification des lois'' \`a l'\'echelle des
galaxies.

De plus le mod\`ele contient l'\'energie noire sous forme d'une constante
cosmologique $\Lambda$. Il offre une sorte d'unification entre l'\'energie
noire et la mati\`ere noire \textit{\`a la} MOND. En cons\'equence de cette
unification on trouve que l'ordre de grandeur naturel de $\Lambda$ doit \^etre
compatible avec celui de l'acc\'el\'eration $a_0$, c'est-\`a-dire que
$\Lambda\sim a_0^2/c^4$, ce qui est en tr\`es bon accord avec les
observations.

Le mod\`ele de mati\`ere noire dipolaire contient donc la physique
souhait\'ee. Son d\'efaut actuel est de ne pas \^etre connect\'e \`a de la
physique microscopique fondamentale (\textit{via} une th\'eorie quantique des
champs). Il est donc moins fondamental que CDM qui serait motiv\'e par exemple
par la super-sym\'etrie. Ce mod\`ele est une description effective, valable
dans un r\'egime de champs gravitationnels faibles, comme \`a la lisi\`ere
d'une galaxie ou dans un univers presque homog\`ene et
isotrope. L'extrapolation du mod\`ele au champ gravitationnel r\'egnant dans
le Syst\`eme Solaire n'est pas enti\`erement r\'esolue. D'un autre c\^ot\'e le
probl\`eme de comment tester (et \'eventuellement falsifier) ce mod\`ele en
cosmologie reste ouvert.

%\appendix

\subsection{Appendice: Champ \'electrique dans un di\'electrique} 

Un di\'electrique est un mat\'eriau isolant, qui ne laisse pas passer les
courants, car tous les \'electrons sont rattach\'es \`a des sites
atomiques. N\'eanmoins, les atomes du di\'electrique r\'eagissent \`a la
pr\'esence d'un champ \'electrique ext\'erieur: le noyau de l'atome charg\'e
positivement se d\'eplace en direction du champ \'electrique, tandis que le
barycentre des charges n\'egatives c'est-\`a-dire le nuage \'electronique se
d\'eplace dans la direction oppos\'ee. On peut mod\'eliser la r\'eponse de
l'atome au champ \'electrique par un dip\^ole \'electrique
$\bm{p}=q\,\bm{\xi}$ qui est une charge $+q$ s\'epar\'ee d'une charge $-q$ par
le vecteur $\bm{\xi}$, et align\'e avec le champ \'electrique. La densit\'e
des dip\^oles nous donne la polarisation $\bm{P}=n\bm{p}$. Le champ cr\'ee par
les dip\^oles se rajoute au champ ext\'erieur (engendr\'e par des charges
ext\'erieures $\sigma_\text{ext}$) et a pour source la densit\'e de charge de
polarisation qui est donn\'ee par la divergence de la polarisation:
$\sigma_\text{pol}=-\bm{\nabla} \!  \cdot \!  \bm{P}$. Ainsi l'\'equation de
Gauss (qui s'\'ecrit normalement $\bm{\nabla} \! \cdot \!  \bm{E} =
\sigma_\text{ext}/\varepsilon_{0}$) devient en pr\'esence du di\'electrique
$\bm{\nabla} \!  \cdot \! \bm{D} = \sigma_\text{ext}$ en utilisant les
conventions habituelles, avec $\bm{D}=\varepsilon_{0}\bm{E}+\bm{P}$. On
introduit un coefficient de susceptibilit\'e \'electrique $\chi_\text{e}$ qui
intervient dans la relation de proportionalit\'e entre la polarisation et le
champ \'electrique: $\bm{P}=\varepsilon_{0}\chi_\text{e}\bm{E}$, ainsi:
$\bm{D}=\varepsilon_{0}(1+\chi_\text{e})\bm{E}$. La susceptibilit\'e est
positive, $\chi_\text{e}>0$, ce qui implique que le champ dans un
di\'electrique est plus faible que dans le vide. C'est l'effet
d'\textit{\'ecrantage} de la charge par les charges de polarisation. Ainsi
garnir l'espace int\'erieur aux plaques d'un condensateur avec un mat\'eriau
di\'electrique diminue l'intensit\'e du champ \'electrique, et donc augmente
la capacit\'e du condensateur pour une tension donn\'ee.


\begin{thebibliography}{}
\bibitem{Mil} M. Milgrom, Astrophys. J. \textbf{270}, 365 (1983).
\bibitem{Be} J.D. Bekenstein, Phys. Rev. D \textbf{70}, 083509 (2004).
\bibitem{Sa} R.H. Sanders, Mon. Not. Roy. Astr. Soc. \textbf{363}, 459
  (2005).
\bibitem{B} L. Blanchet, Class. Quant. Grav. \textbf{24}, 3529 (2007).
\bibitem{BL} L. Blanchet and A. Le Tiec, Phys. Rev. D \textbf{78}, 024031
  (2008); and submitted, arXiv:0901.3114 (2009).
\end{thebibliography}
\end{document}